\documentclass[10pt, preprint]{aastex}

\usepackage{graphicx}
\usepackage{epsfig}
\usepackage{natbib}

\newcommand{\hone}{ H^{(1)}_m }

\newcommand{\rhat}{\hat{\mbox{\boldmath $r$}}}
\newcommand{\zhat}{\hat{\mbox{\boldmath $z$}}}

\begin{document}
\title{Multiple scattering of waves by a pair of gravitationally stratified flux tubes}
\begin{abstract}
We study the near-field coupling of a pair of flux tubes embedded in a gravitationally stratified environment. The mutual induction of the near-field {\it jackets} of the two flux tubes can considerably
alter the scattering properties of the system, resulting in sizable changes in the magnitudes of scattering coefficients and bizarre trends in the phases. The dominant length scale governing the induction zone
turns out to be approximately half the horizontal wave length of the incident mode, a result that fits in quite pleasantly with extant theories of scattering. Higher-$\beta$ flux tubes are more strongly coupled than
weaker ones, a consequence of the greater role that the near-field jacket modes play in the such tubes. We also comment on the importance of incorporating the effects of multiple scattering when studying the 
effects of mode absorption in plage and interpreting related scattering measurements. That the near-field plays such an important role in the scattering process lends encouragement to the eventual goal of observationally 
resolving sub-wavelength features of flux tubes using techniques of helioseismology.
\end{abstract}

\author{Shravan M. Hanasoge$^1$}
\altaffiltext{1}{Visiting academic at Indian Institute of Astrophysics, Bangalore, India}
\affil{Max-Planck-Institut-f\"{u}r-Sonnensystemforschung, 37191 Katlenburg-Lindau, Germany}
\author{Paul S. Cally}
\affil{Centre for Stellar and Planetary Astrophysics, Monash University, Victoria 3800, Australia}
\section{INTRODUCTION}\label{intro.sec}
An outstanding issue in solar physics concerns the accurate constraining of the internal constitution of sunspots. Since we are unable directly image the interior, we study the solar acoustic wave field in and around 
sunspots and attempt
to comprehend these observations through theories of wave interactions. An example of such an effort is the first putative detection of downflows underneath sunspots by \citet{Duvall1996}, who analyzed the solar
wave field using methods of time-distance helioseismology \citep{duvall}.  In the Sun, observations have almost always been more plentiful than theory. Of late, the importance of developing theoretical and computational
methods to aid the interpretation of observations of solar magnetism has risen to the fore. A considerable body of computational work has recently focused on understanding the nature of wave interactions in magnetized
environments \citep[e.g.][]{khomenko06,cameron07,hanasoge_mag}. Complementary to such efforts, we attempt here to develop further the theory of flux tube related multiple scattering, probably
ubiquitously present on the Sun but has, for the most part, been studiously ignored in the past due to the many challenges involved in modeling such interactions.
On another front, it is important to place bounds on the degree of wave absorption and scattering by the plage for it tells us how much
energy is transmitted to the corona and could help explain the complex frequency dependence of acoustic mode linewidths \citep[e.g.][]{bogdan96,hindman08}. Because plage comprises ensembles of compactly
packed thin flux tubes, the interaction with waves possibly lies in the multiple scattering regime. 

In linear theory, when a wave encounters an anomaly of some sort, it is scattered at constant frequency, with the resultant wave field being broadly classified into near- and far-field components. The far-field 
consists of propagating modes that transport some fraction of the incident mode energy away from the scatterer. The near-field is more complex, comprising a number of non-propagating horizontally evanescent waves, that arise when the displacements of the anomaly due to the external wave buffetingü cannot be matched by the set of eigenfunctions of the propagating modes. In the case of stratified flux 
tubes in the Sun, the set of $p$-mode eigenfunctions at constant frequency is an incomplete basis, requiring a supplementary set, an uncountably infinite continuum in fact, of these evanescent near-field functions to complete the basis. 
The problem is exacerbated when the displacement eigenfunctions of the scatterer gain complexity. 

The mathematics required to address the near-field in the case of thin flux tubes was set down by \citet{bogdan95}, 
who termed this non-propagating sheath of waves as an ``acoustic jacket" that envelopes flux concentrations. The uncountable continuum of jacket modes arises as a consequence of an infinitely deep lower boundary, required in order to allow tube  modes to disappear into the solar interior. Unfortunately, numerically computing the near-field jacket in this scenario is all but impossible because of various 
formidable integrals in the equations. In order to 
arrive at an analogous but more tractable problem, \citet{barnes00} introduced an artificial lower boundary, leading to a discrete and countably infinite number of near-field modes. More recently, \citet{hanasoge_birch08}
adopted the model of \citet{barnes00} and estimated the magnitude of the near field jacket and the single scattering by an isolated thin flux tube. They attempted to model the observations of \citet{duvall06}, who 
characterized the scattering of $f$ modes by magnetic flux elements. However, magnetic elements are known to consist of a number of tightly packed flux tubes, all likely within the near-fields of each other. Thus the 
single scattering assumption may not be entirely accurate when interpreting these measurements.
 
The presence of a large body of theory to draw upon makes it easier to proceed towards an understanding of the importance of multiple scattering. When a pair of flux tubes lie in the proximity of each other, their near-fields
communicate and depending on the separation, can dramatically alter the nature of the scatter. Thus accounting for mutual induction of near-fields is rather important when studying plage or other closely spaced
scatterers. \citet{bogdan91}, when considering a pair of flux tubes at a series of separations in an unstratified medium, found evidence for three different scattering regimes that they termed multiple, coherent, and
incoherent. The nomenclature points to differences in the degree of coupling between the two flux tubes, with the incoherent regime no different from isolated body scattering and the multiple regime, substantially
different. Subsequently \citet{keppens94} studied ensembles of flux tubes and found that the degree of absorption was greater in ``spaghetti" models than monoliths, pointing to a way of discerning the differences
between the two.

These efforts have been restricted to unstratified media, mainly due to the considerable mathematical complexity that stratification injects. With the addition of gravity, the external driver and the flux tube
displacement eigenfunctions assume distinct and more complicated forms, possibly destroying a number of resonances hitherto possible in the unstratified case. Furthermore, purely analytical techniques cease to
be of utility, even when considering single scattering, let alone its multiple counterpart. When studying mode mixing due to thin flux tubes, \citet{hanasoge_birch08} were forced to apply a number of methods of
linear algebra in order to reliably estimate scattering coefficients and the near-field jacket. On the other hand, multiple scattering is a bit more of a challenge since one must simultaneously determine the wave fields
of a number of disparate scatterers, all the while keeping in mind that each scattering coefficient is defined according to a coordinate system centered on that specific scatterer. Fortunately, the theoretical 
machinery to study these sorts of problems has been developed decades ago, in the context of fluid mechanics, by e.g., \citet{kagemoto86, linton90}. We adopt these methods in our calculations in order to determine
the degree of mode mixing and scattering from a two-tube system. Please note that in the discussions below, the terms ``near field", ``acoustic jacket" and ``envelope of evanescent modes" are used 
interchangeably. The plan of this paper is as follows. We describe the stratification and basic aspects of the flux tube tube model in $\S$\ref{model.sec}. The method of \citet{kagemoto86} in the 
context of interacting thin flux tubes is discussed in $\S$\ref{meth.sec}. The scattering coefficients derived through the application of these techniques for the two-tube system for different incident modes are presented 
in $\S$\ref{results.sec}. Finally we summarize and conclude in $\S$\ref{conclude.sec}.

 \section{MODEL}\label{model.sec}
The background structure in this calculation, adapted from \citet{bogdan96, hanasoge_birch08}, is an adiabatically stratified, truncated polytrope with index $m=1.5$, gravity ${\bf g} = - 2.775 \times 10^4~{\rm cm~s^{-2}}{\zhat}$, reference pressure $p_0 = 1.21 \times 10^5~{\rm g~cm^{-1}~s^{-2}}$, and reference density $\rho_0 = 2.78 \times 10^{-7}~{\rm g~cm^{-3}}$, such that the pressure and density variations are given by,
\begin{equation}
p(z) = p_0\left(-\frac{z}{z_0}\right)^{m+1},
\label{back.pressure}
\end{equation}
and
\begin{equation}
\rho(z) = \rho_0\left(-\frac{z}{z_0}\right)^{m}.
\label{back.density}
\end{equation}
We utilize a right-handed cylindrical co-ordinate system in our calculations, with coordinates ${\bf x} = (r,\theta,z)$ and corresponding unit vectors $(\rhat,\hat{\mbox{\boldmath $\theta$}},\zhat)$. The photospheric level of the background model is at $z=0$, with the upper boundary placed at a depth of $z_0 = 392~{\rm km}$. Following \citet{barnes00}, we introduce a lower boundary at a depth of 98 Mm. The displacement potential $\Psi({\bf x},t)$ describing the oscillation modes ($t$ is time) is required to enforce zero Lagrangian pressure perturbation boundary conditions at both boundaries. This upper boundary condition is reflective in nature and therefore, possibly not very realistic. The incoming $p_n$-mode, a plane wave, which expanded in cylindrical coordinates \citep[e.g.][]{gizon06} has a displacement eigenfunction, $\Psi_{\rm inc}$, of the form:
\begin{equation}
\Psi_{\rm inc} = \sum_{m=-\infty}^{\infty} i^m J_m(k_n^pr) \Phi_p(\kappa_n^p;s) e^{i [m\theta -\omega t] },\label{incident}
\end{equation}
where,
\begin{equation}
\Phi_p(\kappa_n^p;s) = s^{-1/2-\mu} N_n\left[C^p_n M_{\kappa^p_n,\mu}\left(\frac{s \nu^2}{\kappa^p_n}\right) + M_{\kappa^p_n,-\mu}\left(\frac{s \nu^2}{\kappa^p_n}\right) \right].\label{modes.eigfunc}
\end{equation}
The various symbols in equations~(\ref{incident}) and~(\ref{modes.eigfunc}) are:
\begin{equation}
\mu = \frac{m-1}{2}, ~~~\nu^2 = \frac{m\omega^2 z_0}{g}, ~~~k^p_n = \frac{\nu^2}{2\kappa^p_n z_0},\label{eig.values}
\end{equation}
$\omega$ the angular frequency of oscillation, $s = -z/z_0$, $J_m(w)$, the Bessel function of order $m$ and argument $w$ and $M_{\kappa, \mu}(w)$, the Whittaker function \citep[e.g.][]{whittaker} with indices 
$\kappa, \mu$ and argument $w$. The eigenvalue $\kappa^p_n > 0$ and constant $C^p_n$ characterizing the mode are obtained through the procedure described in appendix A of \citet{hanasoge_birch08}. 
The $n=0$ mode corresponds to the surface gravity or $f$ mode, while $n > 0$ represents the acoustic $p_n$ mode. Note that the lower boundary results in a finite sized box and hence places a restriction on the number of $p$ modes that can fit in this domain. The term $N_n$ is the normalization constant for the mode, defined as
\begin{equation}
N_n = \left[\int_1^{\infty}\left[C^p_n M_{\kappa^p_n,\mu}\left(\frac{\nu^2 s}{\kappa^p_n}\right) +M_{\kappa^p_n,-\mu}\left(\frac{\nu^2 s}{\kappa^p_n}\right) \right]^2 ds \right]^{-1/2}.
\end{equation}

The near-field eigenfunctions are also solutions to the same differential equation that governs the propagating modes:
\begin{equation}
\zeta_n(\kappa^J_n;s) = s^{-1/2 - \mu} \left[C^J_n M_{-i\kappa^J_n,\mu}\left(\frac{i\nu^2}{\kappa^J_n}s\right) + M_{-i\kappa^J_n,-\mu}\left(\frac{i\nu^2}{\kappa^J_n}s\right) \right].\label{jacket.modes}
\end{equation}
As can be seen, the only difference between the form of the propagating and evanescent mode eigenfunctions is the fact that the roots are now imaginary. In order to determine the roots, we perform a high resolution
search for eigenvalues $\kappa^{p,J}_n$ and constants $C^{p,J}_n$; the task is relatively easy for the propagating mode parameters but is reasonably difficult for the jacket modes because the eigenvalues may be
very finely spaced \citep[appendix A of][]{hanasoge_birch08}. Subsequently, tables of the propagating and jacket modes are pre-computed at a range of frequencies. Computations of Whittaker functions over large parameter spaces are fairly non-trivial;
 we use a number of CERNLIB routines to accomplish all these tasks.

\subsection{FLUX TUBE}\label{fluxtube.sec}
Applying the approximations listed in $\S$2 of \citet{bogdan96}, a thin flux tube carrying a magnetic flux of $\Phi_f = 3.88 \times 10^{17} {\rm Mx}$, with constant plasma-$\beta$ everywhere inside the tube is embedded
in the polytrope. The thin flux tube approximation,
\begin{equation}
b(s) \approx \sqrt\frac{8\pi p(s)}{1+\beta}, ~~~~\pi R^2(s) \approx \frac{\Phi_f}{b(s)},\label{tuberadius.eq}
\end{equation}
where $b(s)$ and $R(s)$ are the magnetic field and the radius of the tube at depth $s$, is shown to be accurate to better than a percent in the truncated polytrope situated below $z=-z_0$ or $s=1$ \citep{bogdan96}. Note that the magnetic flux associated with the tube is held constant - different values of $\beta$ therefore result in different $b(s)$ and $R(s)$. The constant-$\beta$ property of the tube follows from the assumption
that thin flux tubes are for all practical purposes in thermal and radiative equilibrium with the external medium.
\subsection{Oscillations of the tube}\label{oscill.tube}
For the cases addressed here, we treat only horizontal kink motions of the flux tube ($\xi_\perp(\omega,s)$), caused by impinging $m=\pm1$ modes  \citep[e.g.][]{bogdan96}. The $m=\pm 1$ modes affect the 
tube oscillations according to the differential equation:
\begin{equation}
\left[\omega^2 z_0 + \frac{2gs}{(1+2\beta)(m+1)}\frac{\partial^2}{\partial s^2}+ \frac{g}{1+2\beta}\frac{\partial}{\partial s}\right]\xi_\perp = \frac{2(1+\beta)}{1+2\beta}\omega^2 z_0\frac{\partial\Psi_{\rm inc}}{\partial x},\label{xiperp}
\end{equation}
where $x = r\cos\theta$. The scattered wave field is computed by matching the horizontal components of the motion of the flux tube to the external oscillation velocities. The manner in which this is accomplished is
 detailed in the following section.


\section{METHOD}\label{meth.sec}
\begin{figure}[!ht]
\begin{centering}
\epsscale{0.5}
\plotone{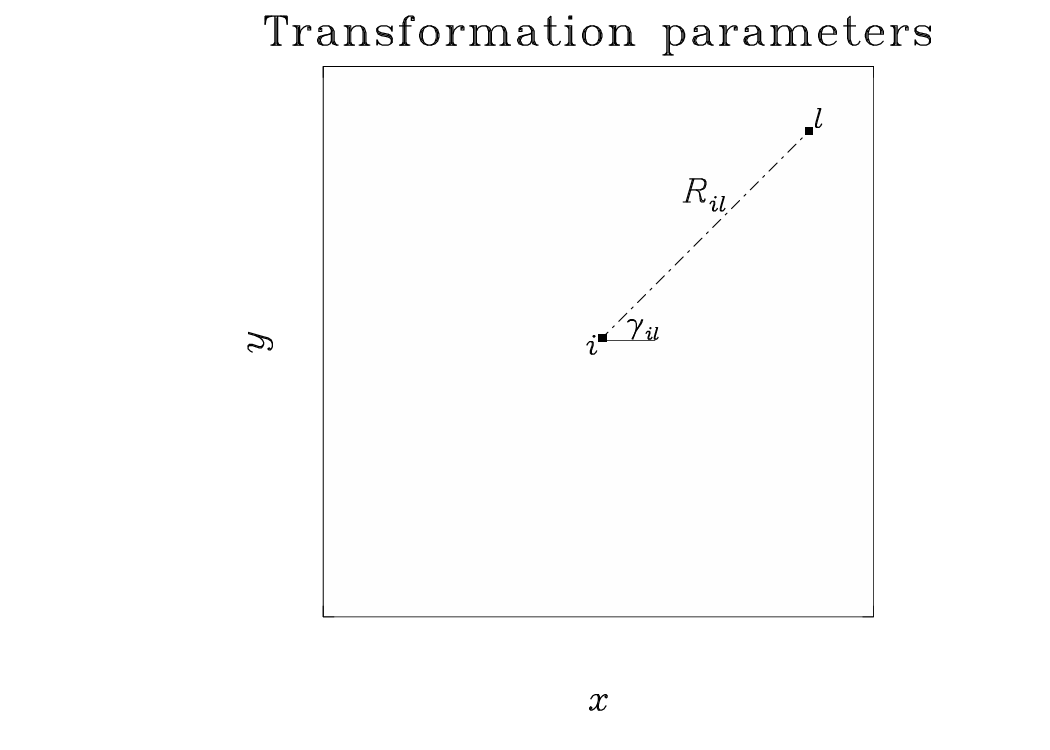}
\caption{The transformation parameters of equations~(\ref{addform.1}) and~(\ref{addform.2}) shown graphically. The tubes $i,l$ are separated by a distance $R_{il}$, with the line joining their centers inclined
at angle $\gamma_{il}$ to the $x$-axis of tube $i$. Note that $\gamma_{li} = \pi - \gamma_{il}$.
\label{transform}}
\end{centering}
\end{figure}

Consider a system of randomly distributed flux tubes. The scattered wave field around tube $i$ with the origin of the coordinate system located at the center of the upper boundary of the tube is given by:
 \begin{equation}
\phi^S_i = -\sum_{m=-1}^{1} \left[  \sum_{n=0}^{n_P}\alpha^i_{mn} \Phi_n(\kappa^p_n; s) \hone(k^p_n r_i) e^{im\theta_i} + \sum_{n=n_P+1}^{N} \alpha^i_{mn}\zeta_n(\kappa^J_n;s) K_m(k^J_n r_i)e^{im\theta_i} \right],
 \end{equation}
where $\zeta_n$ are near-field eigenfunctions, $\Phi_n$ describe the propagating $p$-modes, $H^{(1)}_m(x)$ and $K_m(x)$ are Hankel and K-Bessel functions of order $m$ acting on argument $x$ respectively. 
$n_P$ denotes the number of propagating mode eigenfunctions (a finite number due to the presence of the lower boundary) and the rest corresponding to the evanescent jacket
modes. Also note that the $m$ summation is truncated, since thin flux tube theory applies only to interactions with $|m| \le 1$ waves. Following \citet{kagemoto86}, we write
this in matrix form:
\begin{equation}
\phi^S_i = \sum_n A^T_{in} \Psi^S_{in},\label{element.comp}
\end{equation}
where $A_{in}$ is a vector  (of size $3\times 1$) of scattering coefficients for tube $i$ and mode $n$. The matrix $\Psi_{in}$ (size $3 \times n_z$, with $n_z$ being the number of points in the $z$ grid) contains the partial wave expansions in terms of $\hone, K_m$. In particular, the elements of $A, \Psi$ are:
\begin{eqnarray}
A_{in} &=& -\pmatrix{\alpha^i_{-1n} & \alpha^i_{0n} & \alpha^i_{1n}}^{{\bf T}},\\
(\Psi^S_{in})_{cd} &=& H^{(1)}_{c-2}(k^p_nr_i) \Phi_n(\kappa^p_n;s_d) ~~~~~~~(n \le n_P), \\
(\Psi^S_{in})_{cd} &=& K_{c-2}(k^J_nr_i) \zeta_n(\kappa^J_n;s_d) ~~~~~~~~(n > n_P),
\end{eqnarray}
where $s_d$ is the $d$th point along the $s$ axis, and the indices $c,d$ run from $[1,3]$ and $[1,n_z]$, respectively. As described in the introductory section, the challenge in computing the wave field interactions lies in simultaneously solving for the scattering coefficients of all the 
tubes. Tubes that are placed sufficiently far from each other can only interact via the far-field propagating modes since the evanescent jacket has a spatial decay scale of a wave length or so. Thus the coupling is reasonably
weak since the amplitude of the expanding far-field modal ring falls rapidly with distance from the scatterer. Stronger interactions occur when tubes lie within each other's near-field induction zones, for the evanescent
modes of one influence the tube oscillations on the other and vice-versa. Fairly significant changes in the scattering cross-sections are a consequence of this phenomenon. Thus, to capture this effect, we must compute
the wave field around each tube and project its influence on the neighboring flux elements. Since the wave field around each tube is written in a co-ordinate system centered along its axis, suitable co-ordinate
transformations must be performed. For this purpose, we employ Graf's addition formulae, which transform cylindrical wave functions between co-ordinate systems \citep[e.g.][]{abramowitz}:
\begin{eqnarray}
\hone(k^p_nr_i) e^{im(\theta_i - \gamma_{il})} = \sum_{d=-\infty}^{\infty} H^{(1)}_{m+d}(k^p_n R_{il}) J_d(k^p_nr_l) e^{id(\pi - \theta_l + \gamma_{il})},\label{addform.1}
\end{eqnarray}
\begin{eqnarray}
K_m(k^J_nr_i) e^{im(\theta_i - \gamma_{il})} = \sum_{d=-\infty}^{\infty} K_{m+d}(k^J_n R_{il}) I_d(k^J_nr_l) e^{id(\pi - \theta_l + \gamma_{il})},\label{addform.2}
\end{eqnarray}
where $I_m(x)$ is a Bessel-I function of order $m$ acting on argument $x$. These equations show how to expand an outgoing scattered wave function from tube $i$ in terms of the incident mode waves of tube $l$. 
Here $R_{il}$ is the distance between the centers and $\gamma_{il}$ the angle that the line between the centers of the tubes subtends at the $x$-axis of the tube $i$ (shown graphically in Figure~\ref{transform}).
This relation shows how the scattered wave field 
from $i$ acts as an incident wave on $l$. 
Written in matrix form:
 \begin{equation}
 \Psi^S_{in} = {\bf T}^n_{il}\Psi^I_{ln},\label{tran.eq}
 \end{equation}
 where ${\bf T}$ is the transformation operator that relates the scattered wave field of tube $i$ to the resultant incident field on $l$. We use equations~(\ref{addform.1}) and~(\ref{addform.2}) to build ${\bf T}$; the precise
 distribution of elements is listed in appendix~\ref{matrix.sec}. Note that the index $n$ that appears in equation~(\ref{tran.eq}) denotes a specific modal order, while the superscripts $I, S$ represent the incident and scattered wave expansions. Recalling that equation~(\ref{element.comp}) contains the contributions of various near- and far-field waves and collecting all the $n$'s, we have:
 \begin{equation}
 \phi_i |_l = \sum_n A_{in}^T {\bf T}^n_{il} \Psi^I_{ln}. \label{ionl.eq}
 \end{equation}
Equation~(\ref{ionl.eq}) tells us how much the scattered wave field of tube $i$ acts as an incident field on $l$. Summing up the contributions from the scattered wave fields of all the other flux tubes (except itself, of course)
and the zeroth order incident wave, the total incident wave field at $l$ is given by:
\begin{eqnarray}
\phi^I_l &=& \sum_n\left( \phi_0 |_{ln} +  \sum_{i=1, i\ne l}^{N} A^T_{in} {\bf T}^n_{il} \Psi^I_{ln} \right)\\
&=& \sum_n\left( {a}^T_{ln} + \sum_{i=1, i\ne l}^{N} A^T_{in} {\bf T}^n_{il} \right) \Psi^I_{ln}, \label{total.field.eq}
\end{eqnarray}
and $\phi_0|_{ln} = { a}^T_{ln}\Psi^I_{ln}$, where $a^T_{ln}$  is a vector of coefficients representing the zeroth order incident wave mode of order $n$, acting on tube $l$. The determination of the amplitudes of the 
off-axis incident modes is described in
appendix~\ref{transform.sec}. Note that ${a}_{ln}$ and $A_{ln}$ have the same matrix sizes; the former refers
to the zeroth order incident wave whereas the latter contains the net scattering terms. Finally, to close the equations, we use the diffraction transfer matrix approach of e.g. \citet{kagemoto86} to relate the total incident 
coefficients to the scattered ones. In effect we generate a matrix ${\bf B}_l$ that acts on the incident wave field at $l$ and produces the scattering coefficients. We know the total incident wave field at $l$: it is given by 
equation~(\ref{total.field.eq}). And since $A_l$ is the vector of scattering coefficients as seen by the co-ordinate system centered on $l$, the following relation must hold: 
 \begin{equation}
 {\bf A}_l= {\bf B}_l\phi^I_l,
 \end{equation}
 or,
 \begin{equation}
{\bf A}_l = {\bf B}_l\left( {\bf a}_l + \sum_{i=1, i\ne l}^N  {\bf T}^T_{il} {\bf A}_i\right), \label{scat.eq}
\end{equation}
 where we have the following relations:
 \begin{eqnarray}
 {\bf a}_l &=& \pmatrix{a_{l0} & a_{l1} & a_{l2} &... }^{T},\\
 {\bf A}_l &=& \pmatrix{A_{l0} & A_{l1} & A_{l2} &... }^{T},\\
 {\bf T}_{il} &=& \pmatrix{ T^0_{il} & 0 & 0 & ... &0  \cr 0 &T^1_{il} & 0 & ... &0\cr 0 & 0 &T^2_{il} &... &0\cr ... & ...& ... & ... &... \cr 0 & 0 & ... &0 & T^N_{il}}.\\
 \end{eqnarray} 
 The matrix ${\bf B}_l$ is constructed as follows.  The scattering into all other modes is computed for each incident mode $(m,n)$. This includes both propagating $p_n$ modes and near-field type incident
 waves (the Bessel-I functions).  Thus ${\bf B}_l$ contains a full description of the scattering of any given incident mode $(m,n)$ into any $(m',n')$ for tube $l$. More details are described in appendix~\ref{matrix.sec}.
 
  
Finally, we discuss the means applied to study the simple case of two interacting thin flux tubes $1,2$. Proceeding from equation~(\ref{scat.eq}) we have:
\begin{eqnarray}
{\bf A}_1 &=& {\bf B}_1\left({\bf a}_1 + {\bf T}_{21}^T {\bf A}_2\right),\\
{\bf A}_2 &=& {\bf B}_2\left({\bf a}_2 + {\bf T}_{12}^T {\bf A}_1\right),\\
{\bf A}_1 &=& {\bf B}_1\left({\bf a}_1 + {\bf T}_{21}^T{\bf B}_2\left[{\bf a}_2 + {\bf T}_{12}^T {\bf A}_1\right]\right),\\
\left[{\bf I} - {\bf B}_1{\bf T}_{21}^T{\bf B}_2{\bf T}_{12}^T \right]{\bf A}_1 &=& {\bf B}_1{\bf a}_1 + {\bf B}_1{\bf T}_{21}^T{\bf B}_{2}{\bf a}_2.\label{final.eq}
\end{eqnarray} 
We pre-compute the ${\bf B}$ matrices for a number of different incident mode frequencies and plasma-$\beta$ values. Finally, having constructed the transformation matrices ${\bf T}$ at a series of tube
separations, we apply the standard MATLAB least-squares algorithm (backslash command) to solve equation~(\ref{final.eq}) for ${\bf A}_1$ and then for ${\bf A}_2$. 

\section{RESULTS}\label{results.sec}
\begin{figure}[!ht]
\begin{centering}
\epsscale{0.5}
\plotone{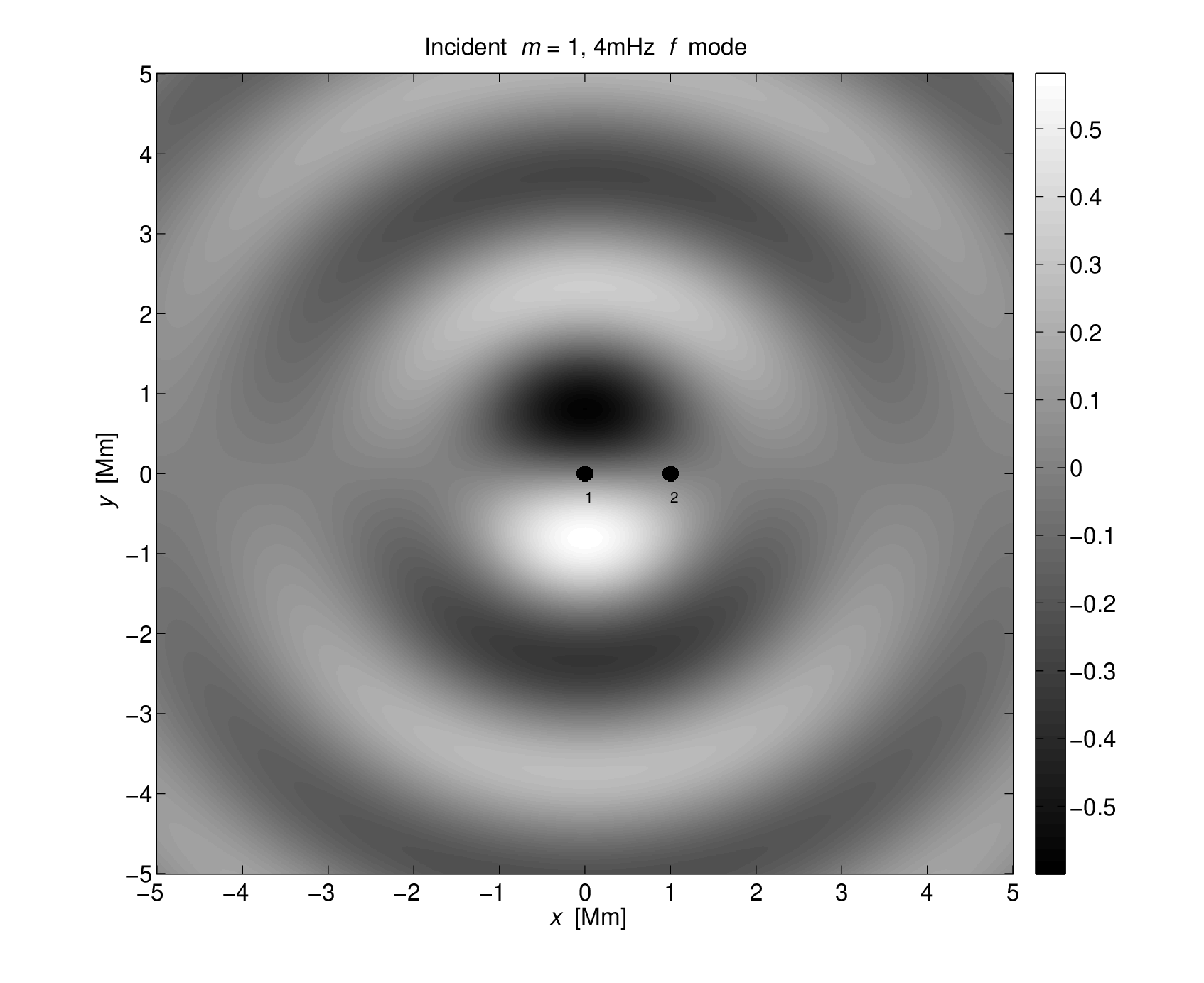}
\caption{Example of an incident $m=1$ wave: $\Psi_{\rm inc} = iJ_1(k^p_0r)e^{i\theta}$. Tube 1 always sees the incident wave, whereas with increasing separation, tube 2 sees less and less of it. Notice that the presence of the second tube kills the $m=\pm1$ symmetry when the tubes are close to each other. The symmetry is restored for wide tube separations, at which point, both tubes are practically isolated bodies.
\label{incwave}}
\end{centering}
\end{figure}

\begin{figure}[!ht]
\begin{centering}
\epsscale{1.}
\plotone{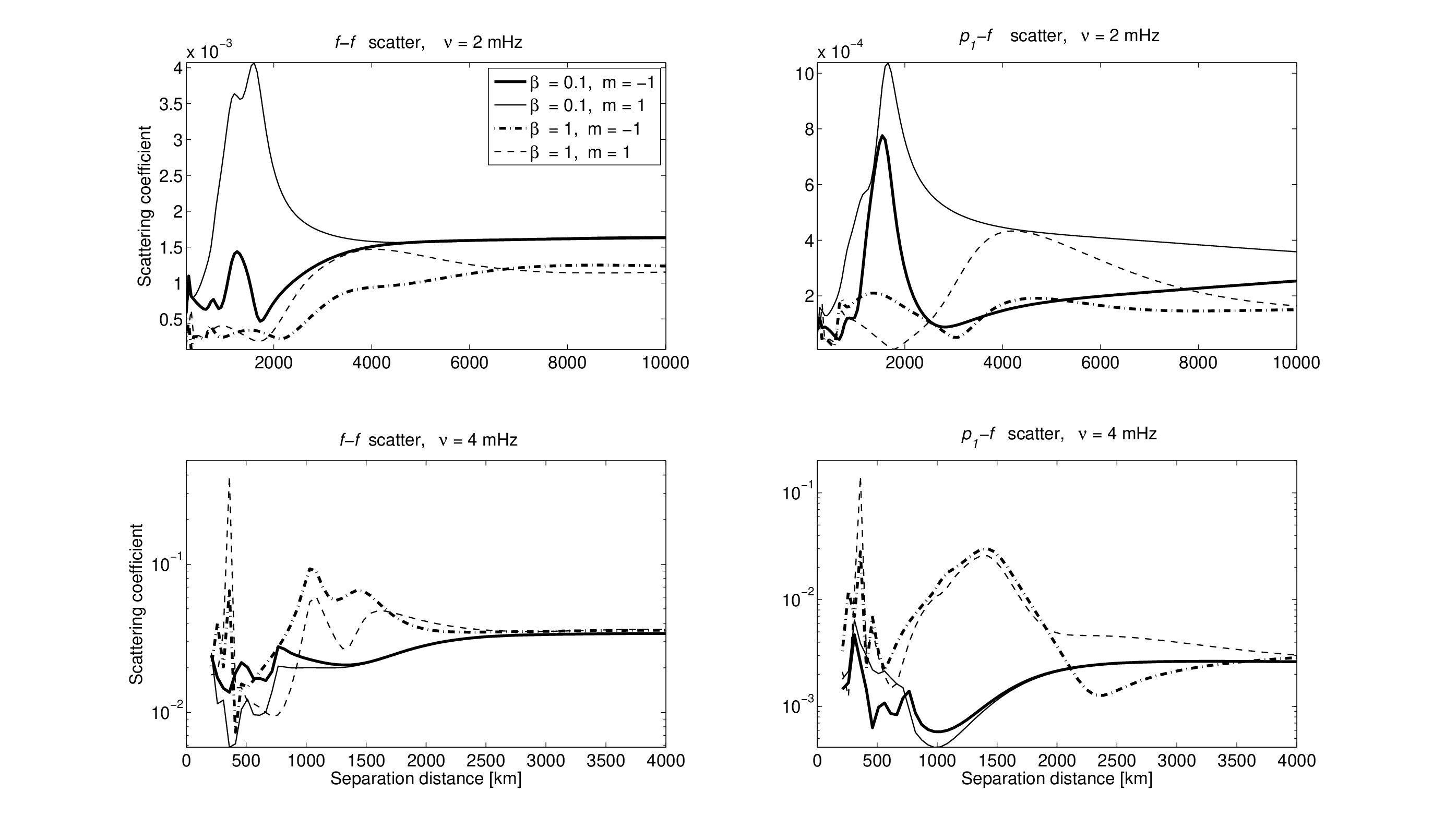}
\caption{Mode-mixing and scattering computed for $f$ and $p_1$ incident waves at two different frequencies and azimuthal orders $m=\pm 1$ for a pair of flux tubes at a number of separations. These are the
coefficients written according to the co-ordinate system centered around tube 1, the tube that encounters the untransformed zeroth order incident mode. Note that the symmetry of the $m=\pm1$ scattering
coefficients are destroyed due to the presence of tube 2; this symmetry is regained when the separation becomes large enough that tube 1 can be considered an isolated body.
Clear signs  of multiple scattering are observed for separations comparable to about half the horizontal wave length at the surface ($\sim \pi/k^p_n$) of the incident waves, which are 11.04 and 2.76 Mm for the $f$ modes and  25.76 and 6.44 Mm for the $p_1$ modes at 2 and 4 mHz respectively. The degree of scatter exhibits a complex behavior at small 
separations and more importantly, shows rather large deviations from the isolated body values. Note also the differences between the $m=\pm 1$ modes; the presence of the second flux tube destroys any symmetries
with respect to these two $m$ waves. Also of interest is the fact that the $\beta=1$ flux tubes couple via the near-field more strongly than the low-$\beta$ tubes; this is presumably because the displacement eigenfunction
exhibits more depth-structure in the former, requiring a greater participation of the near-field modes, and thereby resulting in stronger coupling (see Figure~\ref{eigfuncs}).   
\label{coeffs}}
\end{centering}
\end{figure}

\begin{figure}[!ht]
\begin{centering}
\epsscale{1.}
\plotone{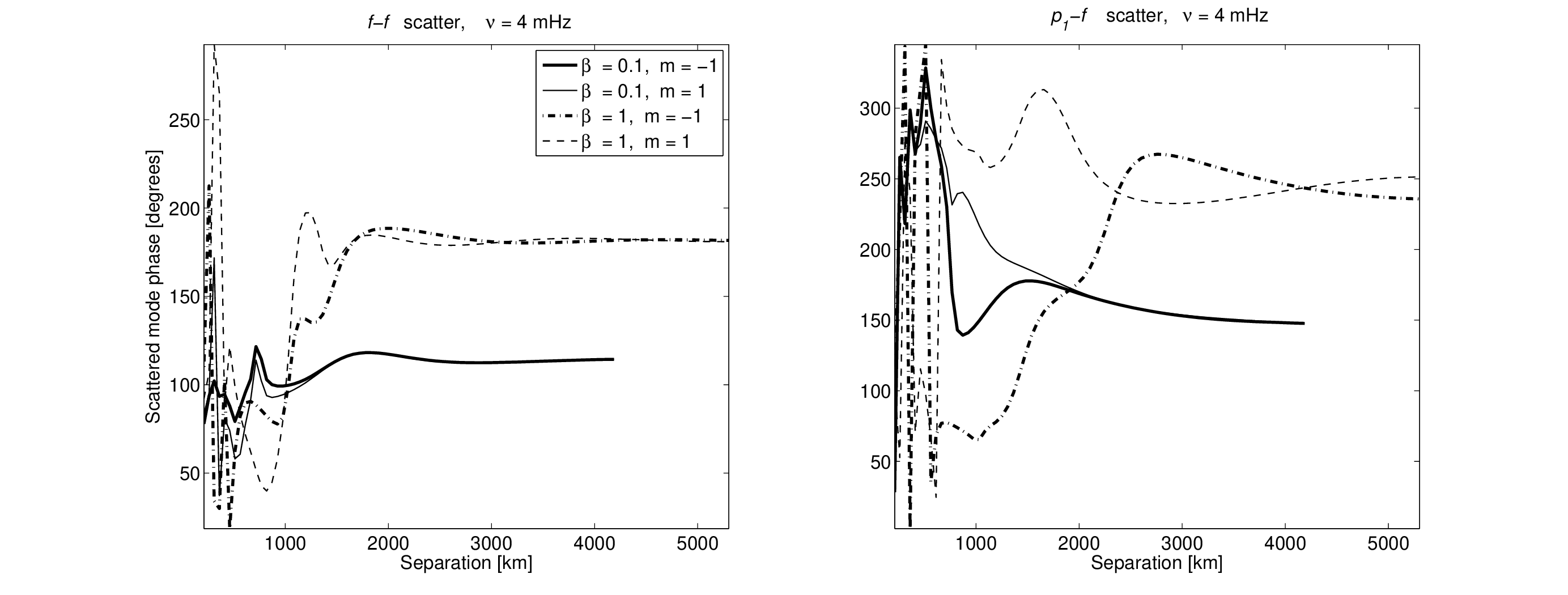}
\caption{Phases of the scattered $f$ mode. The parameter space is similar to that explored in Figure~\ref{coeffs}. No clear pattern is seen, indicating that interpreting the phases
in a multiply scattered wave field is quite a non-trivial affair.
\label{phases}}
\end{centering}
\end{figure}

\begin{figure}[!ht]
\begin{centering}
\epsscale{1.}
\plotone{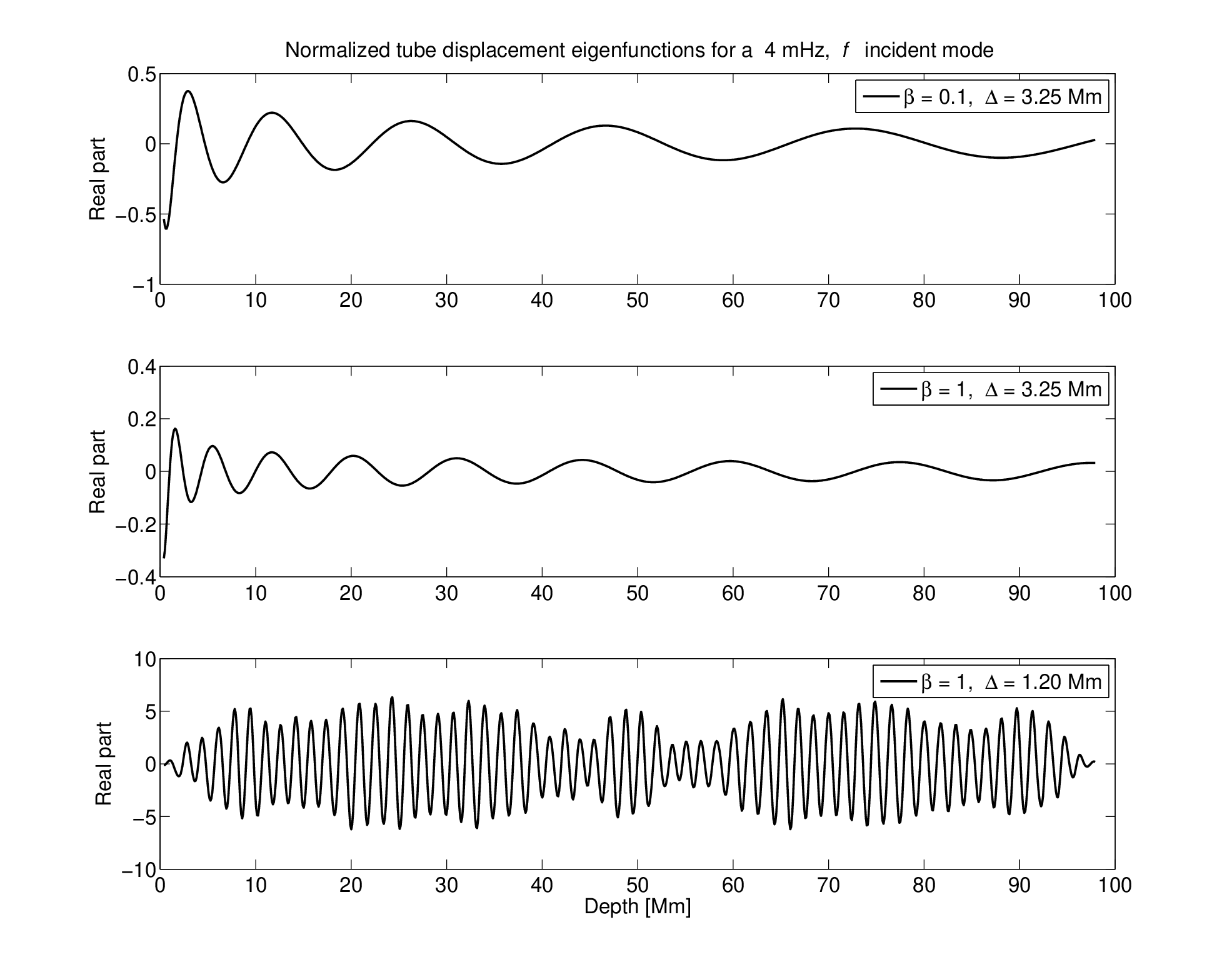}
\caption{Real parts of the normalized tube displacement eigenfunction $(\xi_\perp/k^p_n)$ for flux tube 1 for $\beta =0.1, 1$, and separations, $\Delta = 1.20, 3.25$ Mm. At this frequency, tubes at a separation
distance of 3.25 Mm can be considered§ isolated bodies. The first aspect take note of
 is that the displacement eigenfunction of the 
isolated body  $\beta=0.1$ tube has fewer nodes and therefore less structure in depth than the $\beta=1$ tube (upper two panels). Consequently, a greater near-field participation is required to successfully match
the $\beta=1$ tube displacements with the external wave eigenfunctions, resulting in a more diverse near-field than the $\beta=0.1$ case. Secondly, the lower two panels show how complicated the tube displacement
eigenfunction becomes when the near-fields of the two tubes are in communication. Also, it is important to note that the displacement amplitudes in the strong interaction case are larger by a factor of 10 or more than the
isolated body counterpart. This aspect underscores the very premise of multiple scattering.
\label{eigfuncs}}
\end{centering}
\end{figure}

First we consider the interaction of a pair of identical flux tubes at different separations with incident waves. The angle between the tubes is set to $\gamma_{12} =0$, meaning that the tubes lie on the $x$-axes
of the coordinate systems of each other (see Figure~\ref{transform}).
Undoubtedly there will be large changes in the scattering coefficients at different angles; for now we stick to this simple case, and leave further exploration of the parameter space to future endeavors. A pictorial
representation of an incoming $f$-mode with respect to the flux tubes 1,2 is shown in Figure~\ref{incwave}.
The results of these interactions are shown in Figure~\ref{coeffs}. A number of
conclusions may be drawn: (1) the strongest coupling is between the $f$ mode and the flux tube; there is a rapid fall off with increasing radial order, (2) the length scale of the region occupied by the near-field region 
well approximated by $\pi/k^p_n$, where $2\pi/k^p_n$ is the horizontal wave length of the incident $p_n$ mode, (3) the scattering coefficients attain large values at small flux tube separations; this has
a number of consequences for wave absorption in plage, (4) the phases of the scattered waves exhibit highly unpredictable trends, even more so than the magnitudes of the scattering coefficient (Figure~\ref{phases}),
and (5) higher-$\beta$ flux tubes have more extended jackets than their lower-$\beta$ counterparts; this may be attributed to the fact that when matching
tube displacement eigenfunctions to the external waves, the jacket modes play a more significant role in the former case. The last point is further illustrated by the upper two panels of Figure~\ref{eigfuncs} which show 
the isolated body displacements $\xi_\perp$ of the tubes at plasma-$\beta=0.1,1$. The fact that a larger number of kinks are seen in the higher-$\beta$ tube means a larger number of jacket modes must crowd the
near field region, resulting in stronger coupling between the tubes. The lower two panels of Figure~\ref{eigfuncs} demonstrate the stark differences between the isolated body and near-field coupled tube displacements;
the structure, the number of nodes, and amplitude of the eigenfunction are seen to greatly increase when the jackets of the tubes are able to communicate. This is the very premise of multiple scattering.

\begin{figure}[!ht]
\begin{centering}
\epsscale{1.}
\plotone{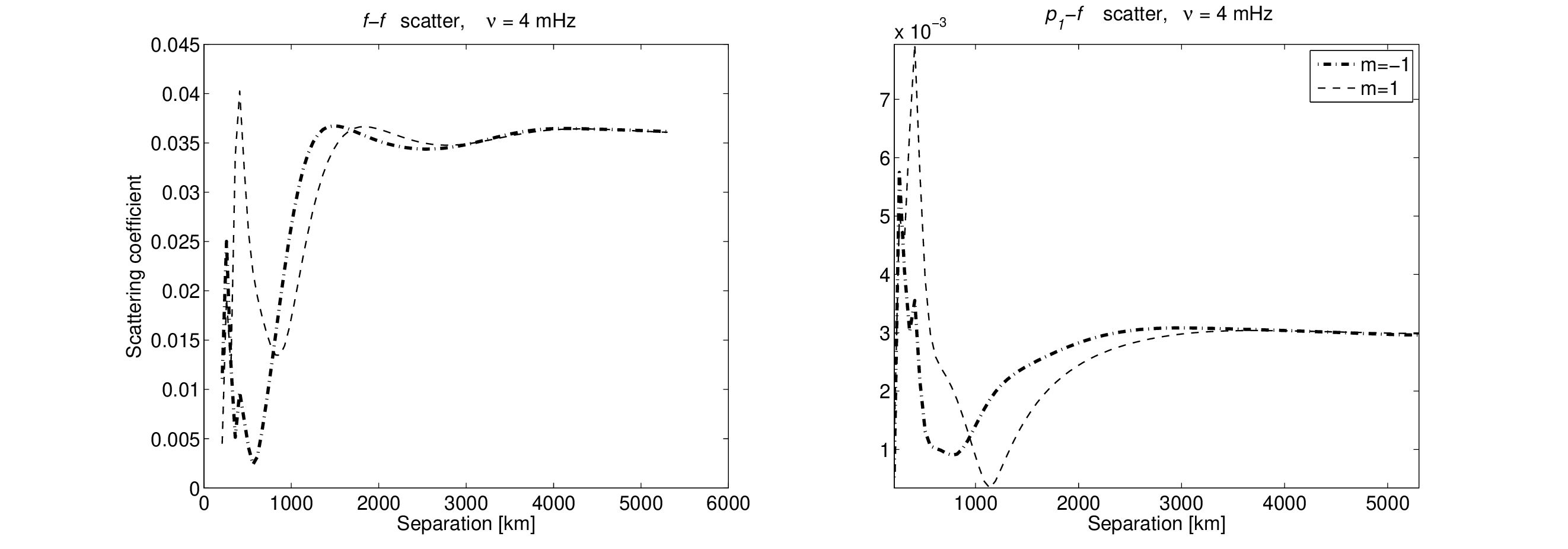}
\caption{Scattering coefficients of the $\beta=1$ tube for an interacting pair of non-identical $\beta=0.1,1$ tubes. The behavior is unremarkable, not greatly differing from that in Figure~\ref{coeffs}. Note that because
the magnetic flux in the tube is held constant, changes in $\beta$ necessarily imply a change in flux tube radius. The incident $f$ and $p_1$ modes have horizontal wavelengths of 2.76 and 6.44 Mm respectively.
\label{difftubes}}
\end{centering}
\end{figure}

\begin{figure}[!ht]
\begin{centering}
\epsscale{1.}
\plotone{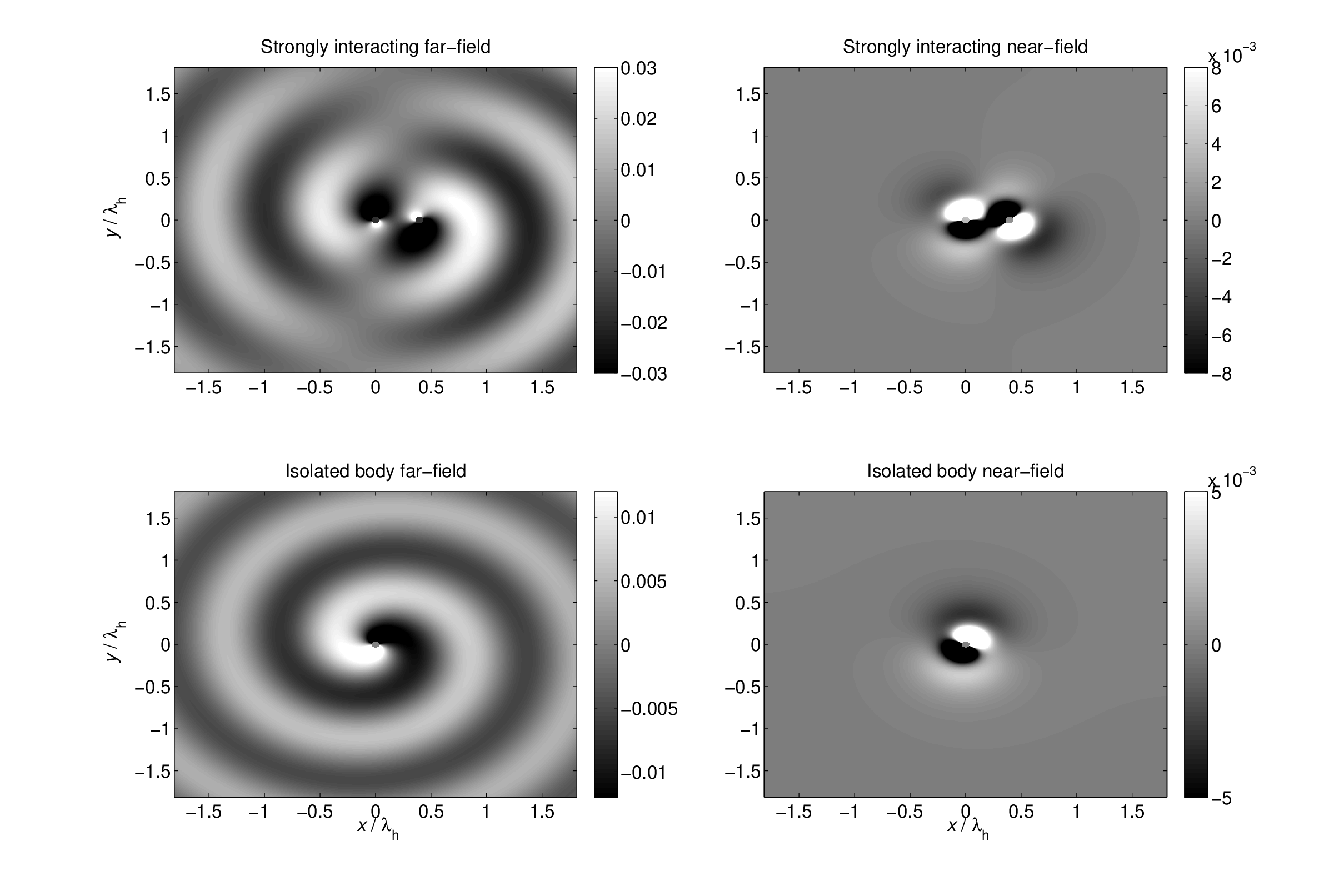}
\caption{Real parts of the near and far scattered wave fields shown pictorially for a pair of $\beta = 1$ flux tubes (upper panels) and the isolated body case (lower panels), both systems lit up by an incident 4 mHz $f$ mode. 
The axes are $x/\lambda_h, y/\lambda_h$, where $\lambda_h$ is the horizontal wave length of the incident $f$ mode, 2.76 Mm. The far-field is seen to spiral out whereas the near-field is bound to within a distance of 
$\lambda_h/2$ of the scatter. Both fields pulsate with a period of 4.16 minutes (4 mHz wave). The tubes are separated by a distance of 1.085 Mm in the upper panels. There may be hope of observationally deducing the
 near-field modes by studying the wave-field auto-correlations and using this information to determine properties of the scatterer.
\label{fields}}
\end{centering}
\end{figure}

We also briefly investigate the interactional behavior of a pair of non-identical flux tubes in Figure~\ref{difftubes}. Because the magnetic flux in each tube is held constant (see $\S$\ref{fluxtube.sec}), the two tubes which 
have differing plasma-$\beta$  of 0.1,1 also are of different radii, commensurate with equation~(\ref{tuberadius.eq}). The scattering coefficient trends are unremarkable, showing differences in structure from those
of Figure~\ref{coeffs} but do not possess any noticeable features. Despite differences in the flux tube geometries, the coupling remains strong and continues to adhere to the half-wavelength rule-of-thumb near-field
dimension. Lastly, in Figure~\ref{fields}, we graph the components of the wave field with the upper panels showing the tubes strongly interacting (a separation distance of 1.085 Mm) while the isolated body case is
displayed in the lower panel. In the strong interaction case, the near-field lobes of the two tubes are seen to be in communication, the scattering in one tube significantly influencing the other through this medium. 
The far-field in both cases are outward spirals, transporting some of the incident mode energy away from the scene of scattering.

\section{CONCLUSIONS}\label{conclude.sec}
The work of \citet{bogdan87} was among the first efforts to characterize and study the multiple scattering of waves by flux tubes. Since then, further activity in this area was restricted to the study of multiple scattering 
by unstratified flux tubes \citep[e.g.][]{bogdan91, keppens94} and subsequently, single scattering by gravitationally stratified tubes \citep[e.g.][]{bogdan96,hanasoge_birch08}. Including stratification when attempting
to solve the full multiple scattering problem introduces manifold difficulties, especially without the right set of techniques. However, much progress in this regard and the availability of the technique of \citet[e.g.][]{kagemoto86}
has allowed us to attempt this problem. 

There has been the general thinking
that not only does gravity inhibit resonant absorption but may in fact prevent strong interactions between closely spaced flux tubes. The reason for this is the disparity in the depth structure between the eigenfunctions of
external modes and the tube displacement; gravity introduces strong structural differences between these two quantities, almost certainly ruling out a strong and direct matching between the two. Thus resonant 
absorption in stratified magnetized environments may be largely ruled out; but what about multiple scattering and spikes in scattering cross sections and phases? Is the theory of single scattering sufficient to 
describe the interaction of waves with clustered magnetic elements in the Sun? Certainly not, as demonstrated in $\S$\ref{results.sec}. Fairly dramatic changes in the scattering coefficients are observed at close
separations, of the order of several hundred kilometers, not unlike the distances between flux tubes in plage. The fact that scatter by a pair non-identical flux tubes also exhibits a similar trend (Figure~\ref{difftubes})
demonstrates the robustness of multiple scattering type interactions. The scattering coefficient is a proxy for the degree of absorption and mode mixing exhibited by the 
system; if nothing else, the jumps in the coefficients point to a loss of coherence of the incident modes, introducing a mechanism for wave damping.

 The coefficients behave in a quirky manner, rather similar to those obtained by \citet{bogdan91}. Thus drawing out larger behavioral properties from these interactions is somewhat difficult
 without a more detailed search of the parameter space. Part of this quirkiness may be attributed to the interference between the sizable numbers of modes competing in the excitation of the flux tubes. Through
 analyses of observations of thousands of small magnetic elements, \citet{duvall06} succeeded in estimating the detailed scattering properties of these features, concluding that the $m=\pm1$ waves couple
 quite strongly with the magnetic tubes. Subsequently, \citet{hanasoge_birch08} modeled these measurements in the single scattering limit in an attempt to constrain properties of the average magnetic element. 
 However, these elements consist of a number of thin flux tubes, the scattering presumably in the strongly interacting regime due to their proximity. Merely with two tubes at a sequence of separations, the scattering
 coefficients and phases display remarkably intricate behavior; extending this to a number of flux tubes at various random locations is arguably a difficult task. 
 
 The near field in Figure~\ref{fields} still remains to be directly detected in observations. One possible route towards this goal is to look for statistically significant auto-correlation signals in the vicinity of these small
 magnetic elements. This follows from the spatially stationary nature of the near-field modes; they merely pulse at the frequency of the incident wave in a thin envelop around the scatterer. However it is unclear
 how to use this information in a productive manner; evidently a greater understanding of these evanescent waves is needed.

 The question relating to sunspot structure, as to whether one can expect
 to be able to helioseismically discern a monolith from a jelly fish, still shows promise, for it would appear that multiple scattering, despite the presence of strong gravitational stratification, still plays a major role. 
 Our preliminary conclusions, derived from Figures~\ref{coeffs} and~\ref{phases} echo those of \citet{keppens94}, who suggested that ensembles of flux tubes can absorb 
 quite effectively, but not cause coherent phasing of the scattered waves.

 This work represents a small first step towards comprehending the action of multiple scattering in stratified magnetized environments. Much work remains to be done in
 terms of characterizing and understanding the interactions of clusters of randomly located flux tubes and the impact on mode linewidths and observations of scattering in plage.

\acknowledgements
The idea for this work occurred over conversations at the CSPA, Monash University, where S. M. H. was a visiting academic in 2008. Much of the computation was performed on the Stanford University
solar group machines; thanks to Phil Scherrer for the use of these resources. Some part of this work was accomplished while S.M.H. was at Stanford, he wishes to acknowledge support
from NASA grant HMI NAS5-02139.
\appendix

 \section{THE {\bf T} and {\bf B} MATRICES}\label{matrix.sec}
The procedure outlined here is adapted from \citet{kagemoto86}. In order to construct ${\bf T}_{il}$, we have the following formula for each $|m,m'| \le 1, n$:
\begin{equation}
({\bf T}_{il})_{cd} = e^{i(m-m')\gamma_{il}} H^{(1)}_{m-m'}(k_n R_{il})
\end{equation}
for $n \le n_{P}$, where $c = 3n +m' + 2$ and $d = 3n +m+2$, and
\begin{equation}
({\bf T}_{il})_{cd} = e^{i(m-m')\gamma_{il}} (-1)^{m'-1}K_{m-m'}(k_n R_{il})
\end{equation}
for $n > n_{P}$, where $c,d$ are as above. The means of constructing ${\bf B}_j$ are as follows. For every tuplet $(|m|\le1,n)$ an incident wave with the appropriate eigenfunction and radial behaviour of unit
amplitude is chosen.  If
$n\le n_P$, the incident mode is propagating and has a $J_m(k^p_n r)$ type horizontal behavior, whereas if $n > n_P$, the horizontal part of the eigenfunction is given by $I_m(k^J_n r)$. For each incident
mode, the scattering into all other modes (at constant frequency and $m$) is computed. Having thus computed the amplitudes of all the scattered waves for each incident mode, we fill up the matrix ${\bf B}_j$ as follows:
\begin{equation}
({\bf B}_j)_{cd} = \beta^m_{n',n},
\end{equation}
where $\beta$ is the scatter into the mode $(m,n')$ by incident wave $(m,n)$. The indices are $c = 3n' + m + 2, d = 3n + m+2$. We only consider $|m| \le 1$ since the tube is insensitive to other
azimuthal orders in this theory.

\section{TRANSFORMATION OF THE INCIDENT WAVE FIELD}\label{transform.sec}
We use Graf's addition formula \citep[e.g.][]{abramowitz}:
 \begin{equation}
J_m(k^p_nr_i) e^{im(\theta_i - \gamma_{ij})} =\sum_{d=-\infty}^{\infty} J_{m+d}(k^p_n R_{ij}) J_d(k^p_nr_j) e^{ip(\pi - \theta_j + \gamma_{ij})}.
 \end{equation}
An $m=0$ wave seen at the axis center of tube $i$ produces the following components at flux tube $l$ whose center is located at distance $R_{il}$ (the $m=[-1,0,1]$ waves):
\begin{equation}
\pmatrix{J_1(k_nR_{il})e^{i\gamma_{il}} & J_0(k_n R_{il}) & J_{-1}(k_nR_{il})e^{-i\gamma_{il}} }.
\end{equation}
Similarly $m=\pm1$ waves seen at tube $i$ produce at $l$:
\begin{eqnarray}
&&\pmatrix{J_2(k_nR_{il})e^{2i\gamma_{il}} & J_1(k_n R_{il})e^{i\gamma_{il}} & J_{0}(k_nR_{il})},\\
&&\pmatrix{J_0(k_nR_{il}) & J_{-1}(k_n R_{il})e^{-i\gamma_{il}} & J_{-2}(k_nR_{il})e^{-2i\gamma_{il}}  },
\end{eqnarray}
respectively. Since in our theory, the tube is insensitive to $|m| > 1$, we do not include these terms in the transformation.


\begin{thebibliography}{19}
\expandafter\ifx\csname natexlab\endcsname\relax\def\natexlab#1{#1}\fi

\bibitem[{{Abramowitz} \& {Stegun}(1964)}]{abramowitz}
{Abramowitz}, M., \& {Stegun}, I.~A. 1964, Handbook of Mathematical Functions
  with Formulas, Graphs, and Mathematical Tables, ninth dover printing, tenth
  gpo printing edn. (New York: Dover)

\bibitem[{{Barnes} \& {Cally}(2000)}]{barnes00}
{Barnes}, G., \& {Cally}, P.~S. 2000, \solphys, 193, 373

\bibitem[{{Bogdan} \& {Cally}(1995)}]{bogdan95}
{Bogdan}, T.~J., \& {Cally}, P.~S. 1995, \apj, 453, 919

\bibitem[{{Bogdan} \& {Fox}(1991)}]{bogdan91}
{Bogdan}, T.~J., \& {Fox}, D.~C. 1991, \apj, 379, 758

\bibitem[{{Bogdan} {et~al.}(1996){Bogdan}, {Hindman}, {Cally}, \&
  {Charbonneau}}]{bogdan96}
{Bogdan}, T.~J., {Hindman}, B.~W., {Cally}, P.~S., \& {Charbonneau}, P. 1996,
  \apj, 465, 406

\bibitem[{{Bogdan} \& {Zweibel}(1987)}]{bogdan87}
{Bogdan}, T.~J., \& {Zweibel}, E.~G. 1987, \apj, 312, 444

\bibitem[{{Cameron} {et~al.}(2007){Cameron}, {Gizon}, \&
  {Daiffallah}}]{cameron07}
{Cameron}, R., {Gizon}, L., \& {Daiffallah}, K. 2007, Astronomische
  Nachrichten, 328, 313

\bibitem[{{Duvall} {et~al.}(2006){Duvall}, {Birch}, \& {Gizon}}]{duvall06}
{Duvall}, Jr., T.~L., {Birch}, A.~C., \& {Gizon}, L. 2006, \apj, 646, 553

\bibitem[{{Duvall} {et~al.}(1993){Duvall}, {Jefferies}, {Harvey}, \&
  {Pomerantz}}]{duvall}
{Duvall}, Jr., T.~L., {Jefferies}, S.~M., {Harvey}, J.~W., \& {Pomerantz},
  M.~A. 1993, \nat, 362, 430

\bibitem[{{Duvall} {et~al.}(1996){Duvall}, {D'Silva}, {Jefferies}, {Harvey}, \&
  {Schou}}]{Duvall1996}
{Duvall}, T.~L.~J., {D'Silva}, S., {Jefferies}, S.~M., {Harvey}, J.~W., \&
  {Schou}, J. 1996, \nat, 379, 235

\bibitem[{{Gizon} {et~al.}(2006){Gizon}, {Hanasoge}, \& {Birch}}]{gizon06}
{Gizon}, L., {Hanasoge}, S.~M., \& {Birch}, A.~C. 2006, \apj, 643, 549

\bibitem[{{Hanasoge}(2008)}]{hanasoge_mag}
{Hanasoge}, S.~M. 2008, \apj, 680, 1457

\bibitem[{{Hanasoge} {et~al.}(2008){Hanasoge}, {Birch}, {Bogdan}, \&
  {Gizon}}]{hanasoge_birch08}
{Hanasoge}, S.~M., {Birch}, A.~C., {Bogdan}, T.~J., \& {Gizon}, L. 2008, \apj,
  680, 774

\bibitem[{{Hindman} \& {Jain}(2008)}]{hindman08}
{Hindman}, B.~W., \& {Jain}, R. 2008, \apj, 677, 769

\bibitem[{{Kagemoto} \& {Yue}(1986)}]{kagemoto86}
{Kagemoto}, H., \& {Yue}, D.~K.~P. 1986, Journal of Fluid Mechanics, 166, 189

\bibitem[{{Keppens} {et~al.}(1994){Keppens}, {Bogdan}, \&
  {Goossens}}]{keppens94}
{Keppens}, R., {Bogdan}, T.~J., \& {Goossens}, M. 1994, \apj, 436, 372

\bibitem[{{Khomenko} \& {Collados}(2006)}]{khomenko06}
{Khomenko}, E., \& {Collados}, M. 2006, \apj, 653, 739

\bibitem[{{Linton} \& {Evans}(1990)}]{linton90}
{Linton}, C.~M., \& {Evans}, D.~V. 1990, Journal of Fluid Mechanics, 215, 549

\bibitem[{{Whittaker} \& {Watson}(1963)}]{whittaker}
{Whittaker}, E.~T., \& {Watson}, G.~N. 1963, {A course of modern analysis}
  (Cambridge: University Press, |c1963, 4th ed.)

\end{thebibliography}

\end{document}